\documentclass[10pt,newlfont,aps,pra,twocolumn,superscriptaddress]{revtex4}
\usepackage{graphicx}
\usepackage{color}
\def\hw{$\hbar\omega_{ho}$}

\def\hs{\hspace{0.5cm}}
\def\ft{\hspace{0.1cm}}
\def\ea{{\it et al.}}
\def\aho{$a_{ho}=\sqrt{\hbar/(2 m\omega_{ho})}$}
\def\hw{$\hbar\omega_{ho}$}
\begin{document}
\title{Long-time averaged dynamics of a Bose-Einstein condensate in a bichromatic optical lattice with external harmonic confinement}
\author{Asaad R. Sakhel}
\affiliation{Department of Physics and Applied Sciences, Faculty of Engineering Technology, Balqa Applied University, 
Amman 11134, Jordan}
\affiliation{Abdus-Salam International Center for Theoretical Physics, Strada Costiera 11, 34151 Trieste, Italy}
\begin{abstract} The dynamics of a Bose-Einstein condensate are examined numerically in the presence of a one-dimensional 
bichromatic optical lattice with external harmonic confinement. The condensate is excited by a focusing red laser. For 
this purpose, the time-dependent Gross Pitaevskii equation is solved using the Crank Nicolson method in real time. Two 
realizations of the optical lattice are considered, one with a rational and the other with an irrational ratio of the 
two constituting wave lengths. For a weak bichromatic optical lattice, the long-time averaged physical observables of the 
condensate respond only very weakly (or not at all) to changes in the secondary optical lattice depth. However, for a much 
larger strength of the latter optical lattice, the response is stronger. It is found that qualitatively there is no 
difference between the dynamics of the condensate resulting from the use of a rational or irrational ratio of the optical 
lattice wavelengths since the external harmonic trap washes it out. It is further found that in the presence of an external 
harmonic trap, the bichromatic optical lattice acts in favor of superflow.
\end{abstract}
\date{\today}
\maketitle

\section{Introduction}

\hs The realization of quasidisorder in a bichromatic optical lattice (BCOL) has hitherto been achieved by the 
superposition of laser beams with an irrational ratio of their wavelengths \cite{Fallani:2007}. As such, the 
resulting quasidisorder and its effects thereof on the properties of a Bose-Einstein condensate (BEC) is of 
significant interest. Indeed, there has been considerable work on the application of this type of OLs 
\cite{Roth:2003,Roscilde:2008,Roux:2008,Cataliotti:2003} and its effects on superfluidity \cite{Gordillo:2015}, 
where a rational approximation to the real irrational wavelength ratio, has been considered. Yet to what extent 
a rational ratio would have on the properties of a BEC in this type of lattice has, to the best of knowledge, not 
been explored, e.g., in terms of the dynamics of the BEC. In the latter case, one talks about a BCOL with quasiperiodicity.

\hs That said, it should be mentioned that earlier, the dynamics of a BEC have been explored in a periodic as 
well as a superlattice \cite{Noort:2007}. In addition, the localization of an expanding BEC has been examined in 
a disordered potential \cite{Palencia:2008}. The latter focused on the regime where the interactions dominate over 
the kinetic energy and disorder; a regime that is considered in this work as well. The dynamics of BECs in quasidisordered 
potentials is also attracting significant attention in a quest for observing Anderson localization in BECs without 
interactions or with repulsive interactions \cite{Palencia:2008}. It is worthy to note, that Adhikari and Salasnich (AS) 
\cite{Adhikari:2009} studied different aspects of localization of a BEC in a 1D quasiperiodic BCOL. They particularly 
investigated the effect of variation of optical amplitudes and wavelengths on the localization without interactions. 
It has been found that a small nonlinearity is capable of destroying the localization of the BEC. Verma \ea\ 
\cite{Verma:2012} studied parametric excitations in an elongated cigar-shaped BEC in a combined 
harmonic trap and a time-dependent OL by using numerical techniques. In other work, Cheng and Adhikari \cite{Cheng:2011} 
studied the localization of a cigar-shaped superfluid Bose-Fermi mixture in a quasiperiodic BCOL for interspecies 
repulsion and attraction; Nath and Roy \cite{Nath:2014} recently provided an exact analytical model for the dynamics 
of a BEC in a BCOL. It has been found that the overlap of two OLs of different depths and incommensurate wavelengths 
results in geometrical frustration of the BCOL. As the depth of the latter rises, the lattice frustration increases 
which allows more inter-site tunneling of the BEC. Therefore, questions of the following type arise: Is it possible 
that in a combined harmonic plus BCOL potential an increased lattice frustration leads to a higher kinetic energy? 
Perhaps a rise in $V_1$ allows more tunneling between the lattice sites? How strong is the influence of the harmonic 
trap as compared to the BCOL in determining the dynamic behavior of the BEC? These are questions that shall be tackled 
in this paper. Of importance is the interplay between the BCOL and the interactions in the degree of localization. What 
is significant in this work here, is that the role of the external harmonic trap is added to this interplay.

\hs Next to this, it should be mentioned that the dynamics of BECs in traps have led to a surge of 
investigations, such as the BECs excited by moving obstacles \cite{Stiessberger:2000,Fujimoto:2010,Fujimoto:2011,
Jackson:2000,Radouani:2004,Caradoc:1999,Caradoc:2000,Neely:2010,Engels:2007,Onofrio:2000,
Madison:2000,Raman:2001,Raman:1999,Madison:2001,Horng:2009}. The obstacle is a potential 
barrier generated by a Gaussian laser beam \cite{Onofrio:2000,Carusotto:2006}, which can 
be repulsive or attractive. So far, mostly repulsive obstacles have been obtained experimentally 
by a blue-detuned laser beam \cite{Neely:2010,Raman:1999,Onofrio:2000,Raman:2001}; here an 
attractive obstacle is considered which is generated by a red-detuned laser.  
\cite{Madison:2000,Madison:2001,Hammes:2002,Garrett:2011,Scherer:2007,Tuchendler:2008,
Stamper-Kurn:1998,Comparat:2006,Jacob:2011,Gustavson:2002,Barrett:2001,Schulz:2007}. The goal
is to investigate whether a combined harmonic plus BCOL trap is able to suppress the effects
arising from exciting the BEC by a stirring agent.

\hs In this work, a comparison is chiefly made between the effects of a BCOL with a rational and irrational 
ratio of the constituting wavelengths on the long-time dynamics of a BEC in the presence of an external harmonic 
trap. An investigation most relevant to the present work has been undertaken by Cataliotti \ea\ \cite{Cataliotti:2003},
who examined the dynamics of a BEC in a combined harmonic and periodic OL potential. There it has been shown 
that the atomic current and phase difference between adjacent lattice sites oscillate at a frequency related to 
the trapping strength and that the frequency of the dipole and quadrupole modes depend on the height $V_0$ of
the OL. The tunneling rate was found to decline with $V_0$. 

\hs The current investigation is for the purpose of examining the effects of a BCOL on the mobility of the bosons 
in a standard mean-field approach. The long-time evolution of the BEC is particularly explored by computing the 
averages of a number of physical observables over extremely long times, the key results being as follows: (1) 
it is demonstrated that the dynamics are not influenced by the differences arising from a quasidisordered or 
quasiperiodic structure of the BCOL as the harmonic trap overcomes them. In this regard, there is no effect 
arising from the lattice frustration; (2) at low primary OL-depths $V_0$, the long-time averaged dynamic physical 
observables of the BEC do not change with an increase of the secondary-OL depth $V_1$. However, for a larger $V_0$ 
beyond a certain value, $V_1$ begins to influence the BEC dynamics. In the latter case, the band structure 
of the system begins to change significantly under the effects of an increased lattice frustration; (3) the 
effects of a secondary OL in the presence of a primary one of a high intensity ($\sim 100\hbar\omega_{ho}$) 
induce modulations in the BEC wavefunction whose effects are manifested through the kinetic term. 

\hs The organization of the present paper is as follows. In Sec.\ref{sec:method} the method is briefly
outlined. In Sec.\ref{sec:results-and-discussion} the results are presented and discussed. In Sec.\ref{sec:conclusions}
the paper concludes with some closing remarks.

\section{Method}\label{sec:method}

\hs The method has been explained earlier in previous work \cite{Sakhel:2011,Sakhel:2013,Sakhel:2015} and the reader
is referred to them for details. Essentially, the time-dependent 1D Gross-Pitaevskii equation (GPE) in units of the
trap \cite{Muruganandam:2009,Sakhel:2013}

\begin{eqnarray}
&&i\frac{\partial\varphi(x;t)}{\partial t}=\,\left[-\frac{\partial^2}{\partial x^2}\,-\,
\frac{\partial^2}{\partial y^2}\,+\,\right.\nonumber\\
&&\left.\frac{\sigma}{4}x^2\,+\,V_{LP}(x;t)\,+\,V_{OL}(x)\,+\,
{\cal G}|\varphi(x;t)|^2 \right]\varphi(x;t), \nonumber \\
\label{eq:GPE}
\end{eqnarray}

is solved numerically via the split-step Crank-Nicolson method \cite{Muruganandam:2009} in real time. The accuracy 
of the CN simulation has been established in Ref.\cite{Adhikari:2009}. In Eq.(\ref{eq:GPE})

\begin{equation}
V_{LP}(x;t)\,=\,A\exp[-\beta(x-vt)^2]
\label{eq:V_LP}
\end{equation}

is the laser potential (LP) with $A$ the depth of the LP, $v$ its velocity, and $\beta$ the parameter describing 
its width. $\varphi(x;t)$ is the wavefunction, $V_{OL}(x)$ the BCOL decribed below, ${\cal G}$ the nonlinear interaction 
parameter given by \cite{Muruganandam:2009}

\begin{equation}
{\cal G}\,=\,2 a_s N \sqrt{2\lambda\kappa}/\ell,
\label{eq:calG}
\end{equation}

with $\lambda$ and $\kappa$ the anisotropy parameters describing the width of the harmonic oscillator ground 
state wavefunction in the $y$ and $z$ directions, $a_s$ the s-wave scattering length, $N$ the number of 
particles, and $\ell=\sqrt{\hbar/(m\omega_{ho})}$ is a length scale so that $a_{ho}=\ell/\sqrt{2}$ is the trap 
length. Is is understood that (\ref{eq:GPE}) was obtained from the 3D GPE after integrating out the $y$ and $z$ 
dependence \cite{Muruganandam:2009}. For the present purpose, we set $\lambda=\kappa=100$ in order to obtain an 
exactly 1D system.  $\varphi$ is the wave function that is normalized to 1. The system is bounded by a box 
potential of size $2L$ so that $-L\le x\le +L$ and in order to enforce the boundary conditions, we set 
$\varphi(x=\pm L)=0$ and $d\varphi(x)/dx|_{x=\pm L}=0$. The units are explained in Sec.\ref{sec:numerics}
below.

\subsection{Bichromatic optical lattice}

\hs The BCOL is generated by 

\begin{equation}
V_{OL}(x)\,=\,V_0 \cos^2(\alpha\pi x)\,+\,V_1 \cos^2(\beta\pi x),
\label{eq:BCOL}
\end{equation}

where $V_0$ is the primary, and $V_1<V_0$ is the secondary OL-depth. The parameters $\alpha$ and $\beta$
determine the periodicity of the OL, that is, whether there is quasiperiodicity or quasidisorder. 
A measure for the strength of quasidisorder can be obtained by computing the standard deviation 
$\delta V=\sqrt{\langle V^2\rangle\,-\,\langle V\rangle^2}$. Hence

\begin{equation}
\langle V\rangle\,=\,\frac{1}{2L}\int_{-L}^{+L} [V_{OL}(x)+V_{ho}(x)]\,dx,
\label{eq:Vav}
\end{equation}

and

\begin{equation}
\langle V^2\rangle\,=\,\frac{1}{2L}\int_{-L}^{+L} [V_{OL}(x)+V_{ho}(x)]^2\,dx,
\label{eq:Vavsq}
\end{equation}

determine a degree of disorder largely influenced by the external harmonic trap.

\subsection{Physical observables}

\hs The time-averaged physical observables that we shall be looking at are the zero-point energy

\begin{equation}
E_{zp}(t)\,=\,\int_{-L}^{+L} dx \left[\frac{\partial\left|\varphi(x;t)\right|}{\partial x}\right]^2,
\label{eq:Ezp}
\end{equation}

the kinetic energy of superflow

\begin{equation}
E_{flow}(t)\,=\,\int_{-L}^{+L} dx \left(\frac{\partial}{\partial x}\phi(x;t)\right)^2 \left|\varphi(x;t)\right|^2,
\label{eq:Eflow}
\end{equation}

$\phi$ being the phase of the BEC, the interaction energy

\begin{equation}
E_{int}(t)\,=\,\frac{{\cal G}}{2}\,\int_{-L}^{+L} dx \left|\varphi(x;t)\right|^4,
\label{eq:Eint}
\end{equation}

the energy due to the external harmonic trap

\begin{equation}
E_{osc}(t)\,=\,\frac{\sigma}{4}\int_{-L}^{+L} dx x^2 \left|\varphi(x;t)\right|^2,
\label{eq:Eosc}
\end{equation}

and the root mean square of the BEC size 

\begin{equation}
R_{rms}(t)\,=\,\left[\langle x^2(t)\rangle\right]^{1/2}\,=\,\int_{-L}^{+L} dx x^2\left|\varphi(x;t)\right|^2.
\label{eq:Rrms}
\end{equation}

We particularly focus on the kinetic energy

\begin{equation}
E_{kin}(t)\,=\,E_{zp}(t)\,+\,E_{flow}(t),
\label{eq:Ekin}
\end{equation}

because the effects of the secondary OL are largely manifested by $E_{kin}(t)$. Its importance lies
in the fact it is a combination of the zero-point energy (quantum pressure) and the kinetic energy
of superflow.

\subsection{Time averaging}

\hs The goal of averaging the physical observables (\ref{eq:Ezp}-\ref{eq:Rrms}) is to study their
long-time evolution. Therefore, the averaging procedure of the form

\begin{equation}
\langle O\rangle\,=\,\frac{1}{T}\,\int_0^T O(t) dt,
\label{eq:Time-Averaging}
\end{equation}

where $O(t)$ is a physical (o)bservable, washes out all the details of the evolution of $O(t)$ with
$t$ and concentrates on the overall performance $\langle O\rangle$ over a very long time $T$. Indeed, 
the latter observables (\ref{eq:Ezp}-\ref{eq:Rrms}) fluctuate with time about a well-defined average 
$\langle O\rangle$. The fluctuations are measured by the variance

\begin{equation}
\delta O\,=\,\left[\langle O^2\rangle\,-\,\langle O\rangle^2\right]^{1/2},
\label{eq:dPO}
\end{equation}

where

\begin{equation}
\langle O^2\rangle\,=\,\frac{1}{T}\int_0^T O^2(t) dt.
\end{equation}

\subsection{Numerics}\label{sec:numerics}

\hs The system considered is a BEC composed of Rb${}^{87}$ atoms at a temperature of $T=0$ K that have an 
s-wave scattering length of $a_s=5.4$ nm \cite{Ruostekoski:2001}. The external harmonic potential has a 
trapping frequency of $\omega_{ho}=2\pi\times 25$ Hz \cite{Ruostekoski:2001} and the box surrounding the 
harmonic trap has a length of $2L=50$. Lengths and energies are in units of the trap 
$a_{ho}=\sqrt{\hbar/(2 m\omega_{ho})}$ and \hw, respectively, and time $t$ is in $1/\omega_{ho}$. Hence, 
$\ell=\sqrt{2}a_{ho}=2.16\mu$m and $a_{ho}=1.53\mu$ m. The interaction strength is set to ${\cal G}=1087.65$ 
[in units of $(\sqrt{2} a_{ho})^{3/2}$], which is in the strongly interacting regime, in order to be in line 
with the Lieb-Liniger parameter in experiments on bosons in a 1D OL \cite{Haller:2010}. The wavefunction 
$\varphi$ is in units of $a_{ho}^{-1}$. From Eq.(\ref{eq:calG}) with $\lambda=\kappa=100$ and the above 
parameters, the number of particles is determined to be $N=1538$. The depth of the LP is chosen $A=-30$, 
width parameter $\beta=4$, and its velocity $v=2$. Using the previous information, $v=1$ in trap units is 
then equal to $2.4\times 10^{-4}$ m/s which is in line with previous experiments \cite{Diener:2002}.

\hs The CN simulations were conducted in the transient regime \cite{Muruganandam:2009,Sakhel:2013,Sakhel:2015},
i.e., after and not including the initialization process, for a substantially long time of $t=10000$ on the 
excellent computational cluster of the Max Planck institute for Physics of Complex Systems in Dresden, Germany. 
In essence, this was a heavy-computational project taking $\sim 2-3$ days of CPU time for each simulation.
The time step used was $t=0.0001$, and the spatial step size was 0.01 therefore yielding 5000 pixels. 

\hs The depths of the BCOL were chosen to that $V_1$ is always less than $V_0$, where $5 \le V_0 \le 100$
(units of \hw). In this work, we set $\alpha=0.4$ and $\beta=1.0$ for a quasiperiodic BCOL on the one hand, 
and $\alpha=0.4$, $\beta=1.3$ for a quasidisordered one on the other hand. The recoil energy in units of the
trap is $E_{rec}\,=\,(\alpha\pi)^2/2$ which for e.g.  $\alpha=0.4$ is 0.7896 so that $V_0\gg E_{rec}$. It must 
be emphasized that the latter ratios of $\beta/\alpha=2.50$ (for $\alpha=0.4$ and $\beta=1.0$), and $\beta/\alpha=3.25$ 
(for $\alpha=0.4$ and $\beta=1.3$), are rational approximations to irrational ratios. That is, had we instead taken e.g. 
$\widetilde{\alpha}=\sqrt{0.161}$ and $\widetilde{\beta}=\sqrt{1.01}$, then $\widetilde{\beta}/\widetilde{\alpha}=2.50654...$ 
is very close to $\beta/\alpha=2.50$ and there would not be much difference between the BCOLs resulting from the 
latter ratios and the dynamics thereof. Is is believed that the above statement is worth investigating experimentally.

\section{Results and discussion}\label{sec:results-and-discussion}

\subsection{Preliminary: strength of quasidisorder}

\hs It is important to examine the change of the standard deviation $\delta V$ with $V_1$ 
to get a sense for the competition between $V_{ho}(x)$ and $V_{OL}(x)$. For this purpose
$\delta V$ is listed in Table \ref{tab:deltaV-various-BCOLS} for various realizations of 
the BCOLs in this work. It can be seen, that for a small $V_0=5$ there is quite a weak 
response of $\delta V$ to changes in $V_1$, whereas for a much larger $V_0=100$ this 
response is more pronounced. The harmonic trap plays a significant role in reducing the
effects of the the quasiperiodic or quasidisordered structure of the BCOL. This in turn is 
manifested in the dynamics of the BEC as shown below.

\begin{table}[t!]
\caption{Standard deviation $\delta V$ [cf. Eqs.(\ref{eq:Vav}) and (\ref{eq:Vavsq})] for
various realizations of the BCOL. From left to right the table lists the primary and secondary
OL depths $V_0$ and $V_1$, respectively, and the standard deviation $\delta V$. The value
of $\alpha$ in Eq.(\ref{eq:BCOL}) is set to 0.4 and three different $\beta$ values are
considered. $V_0$, $V_1$, and $\delta V$ are in units of \hw, whereas $\alpha$ and $\beta$
are in $a_{ho}^{-1}$.}
\begin{tabular}{*5{@{\hspace{0.3cm}} c @{\hspace{0.3cm}}}} \hline\hline
 $V_0$ &  $V_1$ & \multicolumn{3}{c}{$\delta V$ (\hw)} \\ 
 (\hw) &  (\hw) & for $\beta=1.0$ & 1.3  & 1.414... \\ \hline
5.0    &  0.5  &  46.623 & 46.622 & 46.625 \\
       &  1.0  &  46.624 & 46.624 & 46.628 \\
       &  1.5  &  46.626 & 46.625 & 46.632 \\
       &  2.0  &  46.628 & 46.628 & 46.636 \\
       &  2.5  &  46.631 & 46.631 & 46.641 \\
       &  3.0  &  46.635 & 46.634 & 46.647 \\
       &  3.5  &  46.639 & 46.639 & 46.653 \\
       &  4.0  &  46.645 & 46.644 & 46.661 \\ \hline
       &       &         &        &        \\
100.0  & 10.0  &  58.657 & 58.656 & 58.705 \\
       & 20.0  &  58.977 & 58.974 & 59.071 \\
       & 30.0  &  59.506 & 59.501 & 59.645 \\
       & 40.0  &  60.238 & 60.231 & 60.420 \\
       & 50.0  &  61.165 & 61.157 & 61.389 \\
       & 60.0  &  62.280 & 62.270 & 62.542 \\
       & 70.0  &  63.572 & 63.561 & 63.870 \\
       & 80.0  &  65.031 & 65.019 & 65.362 \\
       & 90.0  &  66.646 & 66.632 & 67.007 \\ \hline\hline
\end{tabular}\label{tab:deltaV-various-BCOLS}
\end{table}

\begin{figure}[t!]
\includegraphics[width=7.5cm,bb=60 325 445 679,clip]{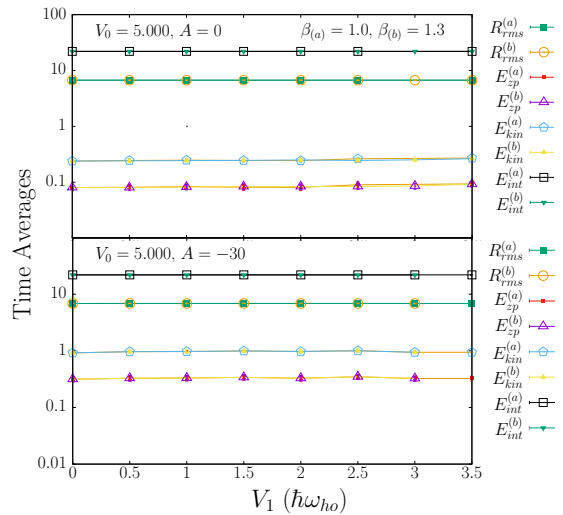}
\caption{(Color online) Time-averaged physical quantities (TAPQs) of a Bose-Einstein condensate in a bichromatic optical lattice 
(OL) with external harmonic confinement as a function of the secondary OL-depth $V_1$. The primary OL has a depth $V_0=5.000$. 
The superscript ($a$) refers to $(\alpha,\beta_{(a)})=(0.4,1.0)$ and ($b$) to $(\alpha,\beta_{(b)})=(0.4,1.3)$. 
In what follows, the left and right angular brackets $\langle ...\rangle$ have been dropped temporarily to allow a clearer
reading of the labels so that the listed observables still represent the time averaged quantities $\langle O\rangle$. Upper 
frame: system is not excited by a laser ($A=0$) and displays the physical quantities $R^{(a)}_{rms}$ (solid squares); 
$R^{(b)}_{rms}$ (open circles); $E^{(a)}_{zp}$ (small solid squares); $E_{zp}^{(b)}$ (open up triangles); $E^{(a)}_{kin}$ 
(open pentagons); $E^{(b)}_{kin}$ (solid up triangles); $E^{(a)}_{int}$ (open squares); $E^{(b)}_{int}$ (solid down triangles). 
Lower frame is as in the upper frame with the same labels; except that that the system is excited by a red laser of depth $A=-30$ 
at a velocity $v=2$. All TAPQs and $V_0$ and $V_1$ are in units of the trap \hw, whereas $R_{rms}$ is in units of \aho.}
\label{fig:plot.timeaveraged.quantities.Vo5.000.several.V1.comparisons.stack}
\end{figure}

\begin{figure}[t!]
\includegraphics[width=7.5cm,bb=60 325 445 679,clip]{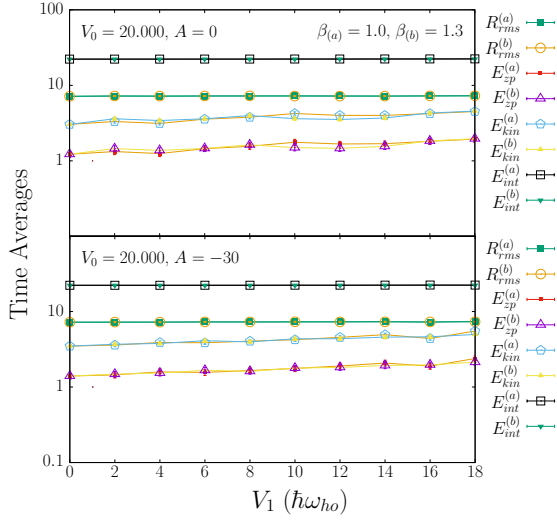}
\caption{(Color online) As in Fig.\ft\ref{fig:plot.timeaveraged.quantities.Vo5.000.several.V1.comparisons.stack}; but for $V_0=20$. All 
energies, $V_0$ and $V_1$ are in units of the trap \hw, whereas $\langle R_{rms}\rangle$ is in units of \aho.}
\label{fig:plot.timeaveraged.quantities.Vo20.000.several.V1.comparisons.stack}
\end{figure}

\begin{figure}[t!]
\includegraphics[width=7.5cm,bb=60 325 445 679,clip]{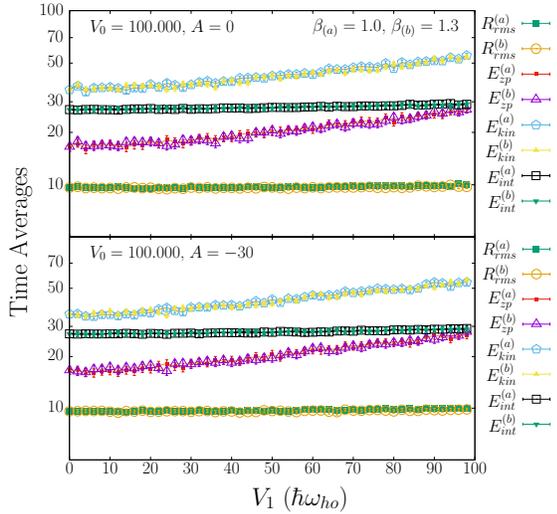}
\caption{(Color online) As in Fig.\ft\ref{fig:plot.timeaveraged.quantities.Vo5.000.several.V1.comparisons.stack}; but for $V_0=100$. 
All energies, $V_0$ and $V_1$ are in units of the trap \hw, whereas $\langle R_{rms}\rangle$ is in units of \aho}
\label{fig:plot.timeaveraged.quantities.Vo100.000.several.V1.comparisons.stack}
\end{figure}

\subsection{Time-averaged physical observables}

\hs The following results display the long-time averaged physical quantities (TAPQs) [Eqs.(\ref{eq:Ezp}-\ref{eq:Ekin})] for various 
realizations of the BCOL. Figure \ft\ref{fig:plot.timeaveraged.quantities.Vo5.000.several.V1.comparisons.stack} shows 
the TAPQs for $V_0=5$ with and without an excitation agent. The TAPQs hardly change with $V_1$ because the interactions 
$\langle E_{int}\rangle$ are dominant and suppress the effects of disorder and kinetic energy $\langle E_{kin}\rangle$. 
The application of a stirring laser (with $A=-30$) causes only $\langle E_{kin}\rangle$ and $\langle E_{zp}\rangle$ to be 
higher than for $A=0$. For $A=-30$ $\langle E_{kin}\rangle \sim 0.94$ and $\langle E_{zp}\rangle\sim 0.32$, whereas for 
$A=0$ $\langle E_{kin}\rangle\sim 0.26$ and $\langle E_{zp}\rangle\sim 0.09$. There is no difference between the BCOL 
with $\beta/\alpha=2.500$ and that with $\beta/\alpha=3.250$. For $V_0=5.000$, the dynamics are therefore largely governed 
by the external harmonic trap, and the system cannot distinguish between the latter two ratios. 

\hs Figure \ft\ref{fig:plot.timeaveraged.quantities.Vo20.000.several.V1.comparisons.stack} is as in 
Fig.\ft\ref{fig:plot.timeaveraged.quantities.Vo5.000.several.V1.comparisons.stack}; but for $V_0=20$. At this 
stage, $\langle E_{zp}\rangle$ and $\langle E_{kin}\rangle$ respond now more to changes in $V_1$ and the difference 
between the two ratios of $\lambda_1/\lambda_2$ has no influence. Again, $\langle R_{rms}\rangle$ and $\langle E_{int}\rangle$ 
remain unaffected. 

\hs At even higher $V_0$ such as 100 in Fig.\ft\ref{fig:plot.timeaveraged.quantities.Vo100.000.several.V1.comparisons.stack}, 
the response of $\langle E_{zp}\rangle$ and $\langle E_{kin}\rangle$ is clearly significant, but the effects due
to the stirring laser are suppressed by the strong BCOL. It is known that the center-of-mass motion of a harmonically 
trapped gas is decoupled from the relative degrees of freedom when the only force acting on the BEC is that due to
the harmonic trap. It is peculiar then, that although the red laser in our work couples to the COM motion of the BEC 
\cite{Sakhel:2015}, the BCOL overcomes the stirring effect of the red laser.

\hs At low $V_0$, the effect of lattice frustration is absent in our systems because the external harmonic trap 
suppresses it. With an increase of the lattice frustation depth, more intersite tunneling of the BEC is allowed 
\cite{Nath:2014} and the kinetic energy rises thereof. Indeed, the secondary OL in the presence of a primary OL 
of a high intensity (e.g. $V_0\sim 100\hbar\omega_{ho}$) induces modulations in the BEC wavefunction whose effects 
are manifested through the kinetic term. The increase of $\langle E_{kin}\rangle$ with $V_1$ at $V_0$ of the order 
of $\sim 100\hbar\omega_{ho}$ indicates that the BCOL is helping to reestablish superflow in the system since $E_{flow}(t)$ 
[Eq.(\ref{eq:Eflow})], included in $E_{kin}(t)$ (\ref{eq:Ekin}), is also rising with $V_1$. The current simulations 
at low $V_0$ are dominated also by the nonlinear interactions. Clearly, there is a competition between the strength 
of the BCOL and the interactions in determining the dynamics of the system at hand. 

\hs An analytic argument can be added by describing the system at hand using a series of Bloch-functions 

\begin{equation}
\Phi_q^{(n)}(x)\,=\,\exp(i q x) u_q^{(n)}(x;t),
\label{eq:Bloch-function}
\end{equation}

where $n$ is the band index, and $q$ the quasimomentum. Hence, the total wavefunction

\begin{equation}
\varphi(x;t)\,=\,\sum_{q,n} \exp(i q x) u_q^{(n)}(x;t),
\label{eq:total-Bloch-wavefunction}
\end{equation}

when substituted into Eq.(\ref{eq:Ezp}), yields that

\begin{eqnarray}
&&E_{zp}(t)\,=\,\int_{-L}^{+L} dx \left|\sum_{q,n} \left\{iq \exp(iqx) u_q^{(n)}(x;t)\,+\,\right.\right.\nonumber\\
&&\left.\left.\exp(iqx)\frac{\partial u_q^{(n)}(x;t)}{\partial x}\right\}\right|^2,
\label{eq:E_zp-Bloch}
\end{eqnarray}

and therefore $E_{zp}\sim q^2$ approximately speaking. As $V_0$ and $V_1$ rise, the total number of
energy bands rises as well yielding a larger $E_{zp}(t)$ according to Eq.(\ref{eq:total-Bloch-wavefunction}).

\begin{figure}[t!]
\includegraphics[width=7.5cm,bb=60 325 445 679,clip]{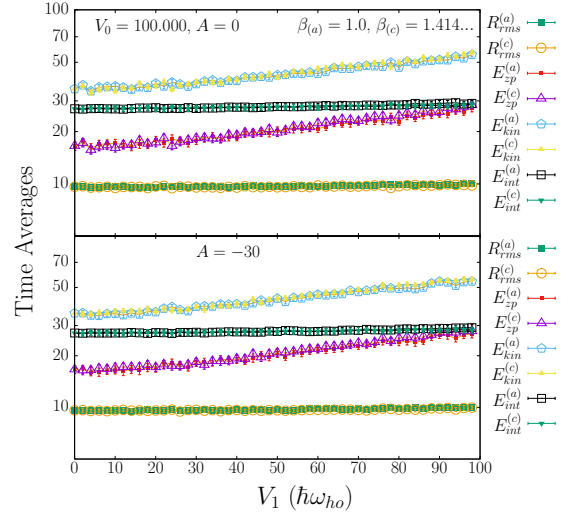}
\caption{(Color online) As in Fig.\ft\ref{fig:plot.timeaveraged.quantities.Vo100.000.several.V1.comparisons.stack}; but for 
$\beta=\beta_{(c)}=1.414213562373095$ instead of $\beta_{(b)}$. All energies, $V_0$, and $V_1$ are in units of the trap \hw, 
whereas $\langle R_{rms}\rangle$ is in units of \aho.}
\label{fig:plot.timeaveraged.quantities.Vo100.000.several.V1.comparisons.freqb1.414.stack}
\end{figure}

\hs What is the effect of using an irrational value for $\beta$ such as $\sqrt{2}=1.4142135623731...$? Will there
be any additional effects thereof? Fig.\ft\ref{fig:plot.timeaveraged.quantities.Vo100.000.several.V1.comparisons.freqb1.414.stack}
is the same as Fig.\ft\ref{fig:plot.timeaveraged.quantities.Vo100.000.several.V1.comparisons.stack}; except that it
uses $\beta=\beta_c=1.414213562373095$ arising from a rational approximation to $\sqrt{2}$. Comparing the two figures, one can
see there is no qualitative difference in the behavior of $\langle E_{kin}\rangle$ and $\langle E_{zp}\rangle$; as
such the external harmonic trap washes out the differences arising from the latter two ratios.

\begin{figure}[t!]
\includegraphics[width=7.5cm,bb=61 276 440 679,clip]{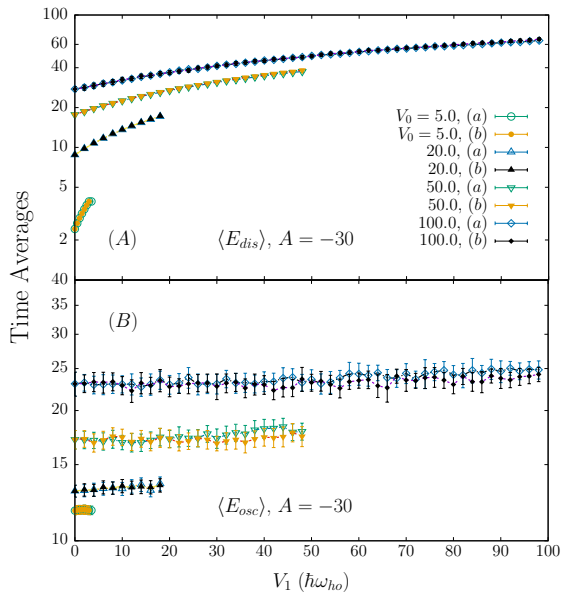}
\caption{(Color online) Upper frame: Time-averaged disorder energy $\langle H_{dis}\rangle$ vs the secondary
OL depth $V_1$ for various values of $V_0$. Open circles: $V_0=5.0$ and $\beta=1.0$; solid circles: 5.0 and 1.3;
open up triangles: 20.0 and 1.0; solid up triangles: 20.0 and 1.3; open down triangles: 50.0 and 1.0; solid down
triangles: 50.0 and 1.3; open diamonds: 100.0 and 1.0; solid diamonds: 100.0 and 1.3. Energies, $A$, $V_0$, and $V_1$
are in units of \hw.}
\label{fig:plot.timeaveraged.Eosc.Edis.several.Vo.and.V1.comparisons.stack}
\end{figure}

\subsection{Disorder vs oscillator energy}

\hs A stronger response to changes in the structure of the BCOL is revealed by the disorder energy

\begin{equation}
E_{dis}\,=\,\int_{-L}^{+L} dx V_{OL}(x;t) \left|\varphi(x;t)\right|^2,
\label{eq:disorder-hamiltonian}
\end{equation}

(which is exclusively connected to the BCOL) as compared to the harmonic oscillator energy $E_{osc}(t)$
given by Eq.(\ref{eq:Eosc}). This is shown in Fig.\ft\ref{fig:plot.timeaveraged.Eosc.Edis.several.Vo.and.V1.comparisons.stack}
for several cases of $V_0$. It can be concluded that the interplay between BCOL and the dynamics of $\varphi(x;t)$ 
is stronger than that between HO and $|\varphi(x;t)|$, as the former yields the stronger response to $V_1$. Further, 
in frame ($A$), the response of $\langle E_{dis}\rangle$ increases as $V_0$ decreases which can be depicted from 
the rate at which $\langle E_{dis}\rangle$ rises with $V_1$.

\subsection{Wavefunction dynamics}

\begin{figure*}[t!]
\includegraphics[width=13.0cm,bb=60 274 595 740,clip]{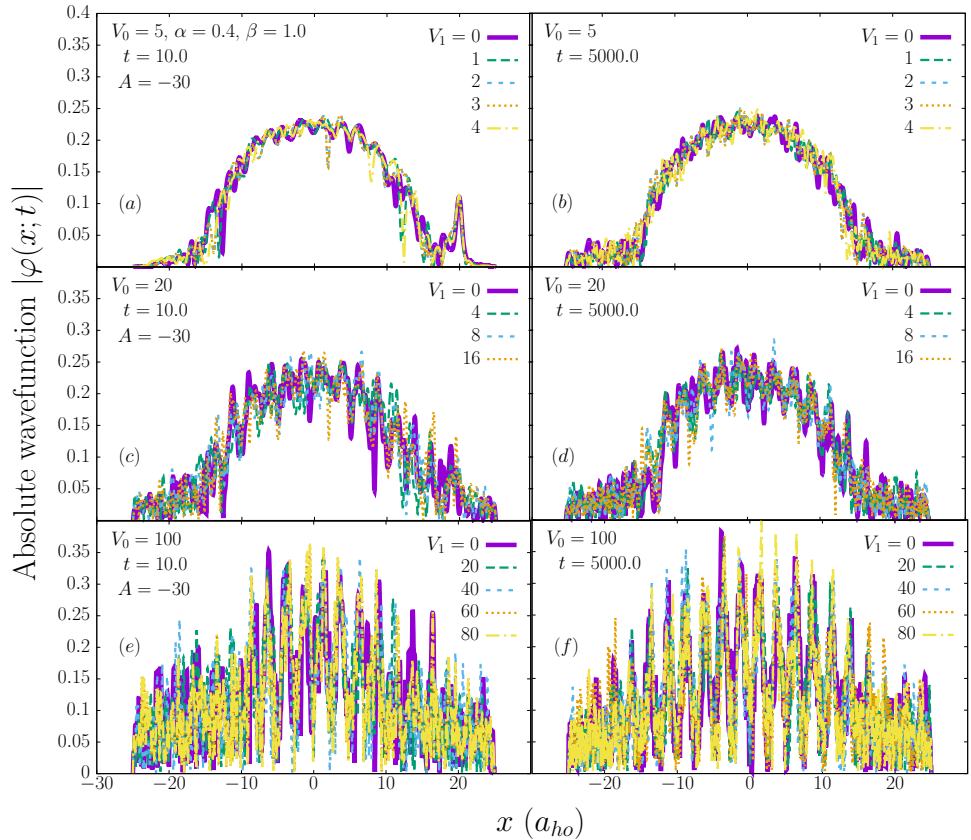}
\caption{(Color online) evolution of $|\varphi(x;t)|$ in a bichromatic optical lattice (BCOL) given by
Eq.(\ref{eq:BCOL}) at different values of the primary and secondary OL depths $V_0$ and $V_1$, respectively. The
periodicity parameters $\alpha$ and $\beta$ are set to 0.4 and 1.0, respectively. The system is excited by a
red laser potential of depth $A=-30$ [see Eq.(\ref{eq:V_LP})]. The left column displays $|\varphi(x;t)|$ at $t=10.0$
when the stirrer is still inside the trap. The right column is at a much later time $t=5000$ when the stirrer has 
long left the trap. Top frames [($a$) at $t=10.0$ and ($b$) at $t=5000$]: $V_0=5$ with $V_1=0$ (solid line); 1.0 
(long-dashed line); 2.0 (short-dashed line); 3.0 (fine-dotted line); and 4.0 (dashed-dotted line). Middle frames
($c$ and $d$): as in the top frames but with $V_0=20$ and $V_1=0$ (solid line); 4.0 (long-dashed line); 8.0
(short-dashed line); 16.0 (fine-dotted line). Bottom frames ($e$ and $f$): $V_0=100$ with $V_1=0$ (solid line);
20.0 (long-dashed line); 40.0 (short-dashed line); 60.0 (fine-dotted line); 80 (dashed-dotted line). $V_0$ and $A$
are in units of \hw, whereas $t$ is in units of $1/\omega_{ho}$.}
\label{fig:plotCNDensityG1087.65A-30B4V2severalVoVbfreqo0.4freqb1.0severaltstack}
\end{figure*}

\hs The evolution of the wave function is of interest because all physical observables (\ref{eq:Ezp}-\ref{eq:Ekin})
are derived from $\varphi(x;t)$. Fig.\ft\ref{fig:plotCNDensityG1087.65A-30B4V2severalVoVbfreqo0.4freqb1.0severaltstack} 
displays $|\varphi(x;t)|$ at different $V_0$ and evolution times. For $V_0=5$, the profile of $|\varphi(x;t)|$ is largely 
determined by the external harmonic trap, although it displays ripples arising from the BCOL. The lattice frustration 
has a significant effect at $V_0=20$ when the BCOL determines the structure of the density profile. Nevertheless,
the increased lattice frustration yields a larger kinetic energy for the superflow as shown in 
Figs.\ft\ref{fig:plot.timeaveraged.quantities.Vo20.000.several.V1.comparisons.stack} and 
\ref{fig:plot.timeaveraged.quantities.Vo100.000.several.V1.comparisons.stack}. For $V_0=100$, the role of the lattice 
frustration is clearly evident as it seems that the bosons tend to localize in the BCOL with increasing $V_0$. However,
what is happening is that the external harmonic trap is trying to localize whereas the BCOL is trying to delocalize
the BEC. The BCOL insreases the local gradients of the wavefunction $\partial |\varphi(x;t)|/\partial x$ and the phase
$\partial\phi(x;t)/\partial x$ necessary for generating the zero-point motion [Eq.(\ref{eq:Ezp})] as well as superflow 
[Eq.(\ref{eq:Eflow})], respectively. In the presence of an external harmonic trap, an increased lattice frustration boosts 
the kinetic energy of the system. The external harmonic trap still plays a role in defining the overall shape of $\varphi(x;t)$, 
yet its dominance has been weakened by the growing influence of the BCOL. The localization observed for $V_0=100$ is similar 
to the classical dynamical transition to a MI state reported earlier by Cataliotti \ea\ \cite{Cataliotti:2003}.

\section{Conclusions}\label{sec:conclusions}

\hs In summary then, the long-time averaged dynamics of a strongly interacting BEC, confined by a combined 
harmonic plus BCOL potential, has been examined. It has been found that the harmonic trap suppresses the 
effects of a BCOL in these dynamics if it is weak $V_0\sim 5$, but for a much stronger BCOL $V_0\sim 100$, 
it begins to compete with the external harmonic trap. Earlier, Verma \ea\ \cite{Verma:2012} have shown that 
there exists a relative competition between the harmonic trap which tries to localize and the optical lattice 
which tries to delocalize the BEC. Indeed, the present work has demonstrated that as $V_0$ and $V_1$ become 
larger (keeping $V_1<V_0$), the increased lattice frustration acts in favor of the superflow by boosting its 
kinetic energy. Qualitatively, there is no difference between the dynamics arising from a BCOL with a rational 
and that with an irrational ratio of the OL wavelengths, since it is found that this is washed out by the 
external harmonic trap. In the future, we shall examine the dynamics under the effects of a changing external 
harmonic trap, i.e., with different values of its strength $\sigma$. This will be accompanied by an examination 
of the corresponding energy band structure of the BCOL.

\acknowledgments

\hs The author wishes to thank the Abdus-Salam International Center for
Theoretical Physics (ICTP) in Trieste, Italy, for a hospitable stay
during which part of this work was carried out. Additional thanks go to the
Max Planck Institute for Physics of Complex Systems (MPIPKS) in Dresden, Germany,
for granting the author access to their excellent computational facilities.
This work has been carried out during the sabbatical leave granted to the author 
Asaad R. Sakhel from Al-Balqa Applied University (BAU) during academic year 2014/2015.

\bibliography{becinquasiperiodiclattice,red_laser}

\end{document}